\documentclass[twocolumn,showpacs,amsmath,amssymb]{revtex4}

\usepackage{graphicx}% Include figure files
\usepackage{dcolumn}% Align table columns on decimal point
\usepackage{bm}% bold math

\def\To{$T_{0}$}
\def\Tc{$T_{c}$}
\def\Tchimax{$T^{\chi}_{max}$}
\def\TchimaxN{$T^{\chi}_{max}(N)$}
\def\Tcmax{$T^{c}_{max}$}

\def\nup{$\nu'$}
\def\cio{$\chi_{0}$}
\def\ket#1{\left\vert #1 \right\rangle}
\begin{document}

\preprint{APS/123-QED}

\title{Monte Carlo analysis of critical phenomenon of the Ising model on memory stabilizer structures}

\author{C. Ricardo Viteri}
\author{Yu Tomita}
\author{Kenneth R. Brown}
\altaffiliation{Author to whom correspondence should be addressed. Electronic mail: ken.brown@chemistry.gatech.edu}
\affiliation{School of Chemistry and Biochemistry and Computational Science and Engineering Division, Georgia Institute of Technology, Atlanta, Georgia 30332, USA}

\date{\today}% It is always \today, today,

\begin{abstract}

We calculate the critical temperature of the Ising model on a set of graphs representing a concatenated three-bit error-correction code. The graphs are derived from the stabilizer formalism used in quantum error correction. The stabilizer for a subspace is defined as the group of Pauli operators whose eigenvalues are +1 on the subspace. The group can be generated by a subset of operators in the stabilizer, and the choice of generators determines the structure of the graph. The Wolff algorithm, together with the histogram method and finite-size scaling, is used to calculate both the critical temperature and the critical exponents of each structure. The simulations show that the choice of stabilizer generators, both the number and the geometry, has a large effect on the critical temperature. 
\end{abstract}

\maketitle

\section{\label{sec:Int}Introduction}

A bit of information can be stored in any physical system with two distinct states. For the physical system to be a reliable memory, the states of the system must be robust against external fluctuations. The classic example is a ferromagnet. Below a critical temperature, the size of the average magnetization is robust against changes in external magnetic field and temperature-driven spin fluctuations.  

A quantum bit (qubit) of information can be stored in any physical system with two orthogonal quantum states. A goal of quantum information is to engineer a system that can reliably store the state of the qubit in the presence of environment-induced fluctuations. A number of approaches have been proposed from quantum error-correction~\cite{Gottesman2009} to passive protection of the information through symmetries~\cite{LidarWhaley} or energetics~\cite{Bacon2001,Kitaev2003,Doucot2005,Bacon2006,Bacon2008}.

The possibility of engineering the quantum equivalent of the magnetic hard drive is quite attractive. The premise is that a macroscopic number of qubits with multi-qubit interactions could create a single stable qubit memory. It is widely suspected that Kitaev's toric code on a four dimensional lattice would achieve this task ~\cite{Kitaev2003, Dennis2002}. In lower-dimensions, the answer is unclear. Bravyi and Terhal have recently shown that for interactions based on stabilizer codes, there is no two-dimensional self-correcting quantum memory~\cite{Bravyi2009}. They make the reasonable physical assumption that the number of qubits involved in the interactions does not grow with the size of the lattice. In the case of self-correcting memories based on concatenated codes the number of qubits involved in each interaction does grow with the lattice size ~\cite{Bacon2008}. 

Here we consider the classical concatenated triple-modular redundancy code in the formalism of quantum stabilizers. The standard choice of generators for this code leads to interactions that grow with the system size. Choosing a different set of generators yields interactions that are only between two-bits and equivalent to an Ising model. In this paper, Monte Carlo simulations are used to study the critical behavior of the error-correction inspired structures shown in Fig.~\ref{fig:structures} within the framework of a ferromagnetic Ising model. The calculations use the Wolff algorithm~\cite{Wolff1989} together with the histogram method~\cite{Ferrenberg1988, Ferrenberg1989} and finite-size scaling~\cite{Fisher1972}. These high resolution Monte Carlo techniques have been utilized successfully to study the critical phenomena of many different model Hamiltonian systems such as the 3D Ising ferromagnet~\cite{Ferrenberg1991}, Heisenberg lattice~\cite{Chen1993}, XY models~\cite{Kim2006,Li1989}, dilute Ising magnet~\cite{Wang1990}, Potts models~\cite{Bonfim1991}, and Sierpinski fractals of dimensions, $d$, between one and two~\cite{Monceau1998, Monceau2001, Monceau2003} and between two and three~\cite{Monceau2002}. We study how the choice of generators changes the Hamiltonian and affects the magnitude of the critical temperature. Critical behavior is characterized by the set of critical exponents ($\alpha$, $\beta$, $\gamma$) and Wolff dynamical critical exponents are calculated for each structure.

The results presented here show that structures with low dimensionality and two-body interactions preserve one bit of information. It suggests two new directions for examining self-correcting quantum memories: 1) choosing non-standard stabilizer generators to minimize the multi-qubit interactions and 2) to examine stabilizer codes on fractional dimensional geometries.

\section{\label{sec:ModMeth}Model and Methods}
\subsection{Stabilizer to Structure}
We consider the familiar classical code of triple modular redundancy. Each bit, $x$, is encoded into a logical bit, $x^L$, consisting of three bits of equal value, i.e. $0^L=000$ and $1^L=111$. If an error occurs on a single bit, majority vote can be used to determine the value of the logical bit; two errors would not be corrected. One way to protect against higher errors is to concatenate the code recursively. At each level of concatenation $k$, the logical bit consists of three bits of level k-1, e.g. a 0 at level $k=2$ is defined as $0^{k=2}=0^{k=1}0^{k=1}0^{k=1}=000000000$. Correction works by majority vote at the lowest level first and then working up. Each level of concatenation $k$ can always correct a maximum of $2^{k}-1$ errors on the physical bits.

The basic idea of the stabilizer formalism is that a quantum state or subspace can be described by the operators that have +1 eigenvalue on that space ~\cite{Nielsen_book}. The stabilizer formalism is particularly useful for describing quantum error correcting codes. Classical error correcting codes represent a subset of quantum error correcting codes that only protect against classical bit-flip errors and not the phase errors. The triple modular redundancy code is a textbook example for introducing the idea of stabilizer error correcting codes ~\cite{Nielsen_book}. Below we review the case of this concatenated code and show how it translates directly onto the Ising model. 

Following standard quantum computation notation, the states of the $j^{th}$ spin are represented as $\ket{0}_j$ and $\ket{1}_j$, $Z_j$ is the Pauli-$z$ operator on the $j^{th}$ spin ($Z_j\ket{0}_j=\ket{0}_j$ and $Z_j\ket{1}_j=-\ket{1}_j$), and $X_j$ is the Pauli-$x$ operator on the $j^{th}$ spin (flips the $j^{th}$ bit). For level-1 encoded bits $\ket{0}^{k=1}=\ket{0}_1\ket{0}_2\ket{0}_3=\ket{000}$ and $\ket{1}^{k=1}=\ket{1}_1\ket{1}_2\ket{1}_3=\ket{111}$. The encoded $Z$ operator is defined as $Z^{k}=Z^{k-1}_1Z^{k-1}_2Z^{k-1}_3$ and for level 1 is $Z^{k=1}=Z_1Z_2Z_3$.

For stabilizer codes, the stabilizer is defined as all the products of Pauli operators that act trivially on the code space. For the three-bit code, the stabilizer consists of  four operators $S^{k=1}=\{I,Z_1Z_2,Z_1Z_3,Z_2Z_3\}$ where $I$ is the identity. The stabilizer can also be defined by the generators of the group, $S^{k=1}=<Z_1Z_2,Z_2Z_3>$.

In the case of level-1 triple modular redundancy, the freedom in the minimal generators is trivial $S^{k=1}= <Z_1Z_2,Z_2Z_3> = <Z_1Z_2,Z_1Z_3>$. Equating the sum of the generators with a Hamiltonian, $H=-J(Z_1Z_2+Z_2Z_3)$, where $J$ is the coupling strength, yields the Ising interaction between three spins in a line. If the full stabilizer is used, then $H=-J(Z_1Z_2+Z_2Z_3+Z_1Z_3+I)$, which corresponds to Ising interactions between spins on a triangle with an energy offset due to the identity. In all cases, the codespace is the degenerate ground state of the Hamiltonian.

We are interested in the thermodynamic limit and whether the energy gap between the ground state and the transition state will preserve the information. Our goal is to study how the choice of generators affects the critical temperature of the ferromagnetic phase transition. At higher levels the choice of generators is non-trivial. As an example, consider k=2. The standard choice of generators is $S^{k=2}=<Z_1Z_2,$ $Z_2Z_3,$ $Z_4Z_5,$ $Z_5Z_6,$ $Z_7Z_8,$ $Z_8Z_9,$ $Z_1Z_2Z_3Z_4Z_5Z_6,$ $Z_4Z_5Z_6Z_7Z_8Z_9>$. Notice the six bit operators can be written as products of $Z^{k=1}$ operators, $Z_1^{k=1}Z_2^{k=1}=Z_1Z_2Z_3Z_4Z_5Z_6$. If the generators are used to define a Hamiltonian, then increasing $k$ leads to many-body operators that act on $2\times3^{k-1}$ spins at once. This exponential increase in the many-body nature of the Hamiltonian makes the physical construction of such a system unlikely. 

\begin{figure}
\centering
\includegraphics[width=0.45\textwidth]{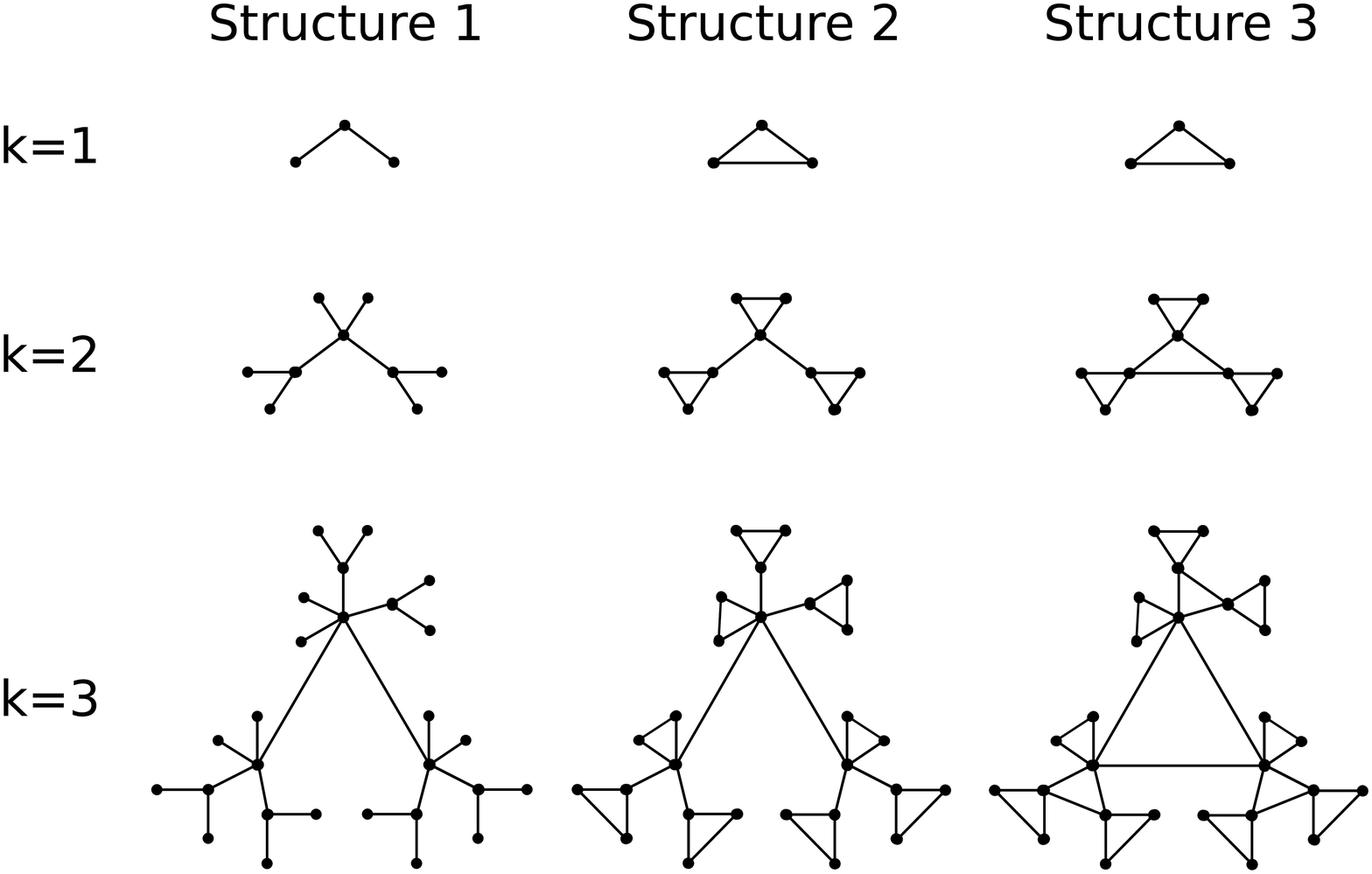}
\caption{\label{fig:structures} Memory stabilizer structures generated by two body interactions. Circles are spin sites (qubits) and the lines show pairs of interacting spins (generators). The interaction strength $J$ is constant (see Eq.~\ref{eq:IsingHamiltonian}). The total number of bits increases with concatenation level, $k$, as $3^{k}$. Only the first three levels are shown.}
\end{figure}

In contrast, we can choose a set of pairwise Ising interactions that generate the same group. A natural choice would be the line $S^{k=2}=<Z_1Z_2,$ $Z_2Z_3,$ $Z_3Z_4,$ $Z_4Z_5,$ $Z_5Z_6,$ $Z_6Z_7,$ $Z_7Z_8,$ $Z_8Z_9>$ but it is well-known that the 1D Ising model does not have a phase transition at finite temperature. We instead consider a model where the encoded bits are Ising coupled middle to middle not end to end: $S^{k=2}=<Z_1Z_2,$ $Z_2Z_3,$ $Z_2Z_5,$ $Z_4Z_5,$ $Z_5Z_6,$ $Z_5Z_8,$ $Z_7Z_8,$ $Z_8Z_9>$. This describes the tree labeled Structure 1 in Fig. \ref{fig:structures}. Structure 2 and 3 are modifications that include loops in the structure. The loops are equivalent to choosing a non-minimal set of generators.

For each structure, the total number of bits, $N$, increases with concatenation level, $k$, as $N=3^{k}$. The set of generators that corresponds to the structure defines the Hamiltonian,

\begin{equation}
H=-J\sum_{\langle i,j \rangle}Z_{i}Z_{j},
\label{eq:IsingHamiltonian}
\end {equation}
where the sum is over nearest-neighbors. $J$ sets the energy scale for the problem and temperature is measured in units of $J/k_B$. The stability of the structure as measured by the phase transition temperature, \Tc, depends on the energy-barrier that separates the two ground states and the number of pathways that traverse the barrier. For these complicated structures \Tc\ must be calculated numerically.

\begin{table}
\caption{\label{tab:structures}Average coordination and number of generator elements per spin site of the three memory stabilizer graphs described in Fig.~\ref{fig:structures} at the $k \rightarrow \infty$ limit.}
\begin{ruledtabular}
\begin{tabular}{cccc}
Structure & Coordination number & Generators per spin site \\ \hline
1	& 2	& 1 \\	 	 
2	& 2$\frac{2}{3}$ & 1$\frac{1}{3}$ \\
3	& 3 & 1$\frac{1}{2}$ \\
\end{tabular}
\end{ruledtabular}
\end{table}

\subsection{Calculating thermodynamic properties}

In the canonical ensemble, the thermodynamic average of an operator $A$, $\langle A \rangle_T$, is given by $Tr[A\exp(-H/k_{B}T)]/Z(T)$,  where $Z(T)=Tr[\exp(-H/k_{B}T)]$ is the partition function. In practice this average cannot be exactly calculated in the limit of large $N$ and numerical approximations are required. Our focus is on calculating the following thermodynamic properties: the average energy, $\langle E \rangle_{T}$, and the average absolute magnetization, $\langle M \rangle_{T}$. At \Tc\ the fluctuations of these quantities diverge for an infinite system, and \Tc\ can be determined by examining the specific heat capacity $c(N,T)$ and the zero field magnetic susceptibility $\chi(N,T)$ as given by

\begin{equation}
c\left(N,T\right)=\dfrac{1}{N}\dfrac{\langle E^{2} \rangle_{T}-\langle E \rangle^{2}_{T}}{k_{B}T^{2}},
\label{eq:hc}
\end {equation}

\begin{equation}
\chi\left(N,T\right)=\dfrac{1}{N}\dfrac{\langle M^{2} \rangle_{T}-\langle M \rangle^{2}_{T}}{k_{B}T}.
\label{eq:magsus}
\end {equation}

\noindent A direct measurement of the degree of preservation of the information can be read from the average magnetization per spin site, defined as $m(N,T)=(1/N)\langle M \rangle_{T}$. Below \Tc\ the system develops spontaneous magnetization and the single order parameter $m$ approaches the value of 1.

Without analytical expressions for $\langle E \rangle_T$ and $\langle M \rangle_T$ , there are three computational challenges: estimating thermodynamic averages for specific values of $T$, determining \Tc\ from the evaluation of thermodynamic averages at a finite set of $T$, and extending our results to the limit of large $N$. We solve each problem using well established numerical techniques for statistical mechanics. 

Reliable studies of thermodynamic averages near critical temperatures require simulations of very large systems. Two possible simulation methods are Metropolis Monte Carlo and the Wolff algorithm ~\cite{Wolff1989,Newman_book}. A Metropolis Monte Carlo step updates the configuration of spins by flipping randomly (one at a time) $N$ chosen spins. Groups of adjacent spins tend to point in the same direction near the critical region, giving rise to correlations in the system. The linear size of these clusters (correlation length, $\xi$) diverges at the critical temperature and successive configurations of spins are generally strongly correlated. The efficiency of the Metropolis algorithm is hindered by the increasing of the number of steps needed to obtain uncorrelated spin configurations~\cite{Monceau2003}. One way to overcome this critical slowing down is by choosing the Wolff algorithm, in which a cluster is identified and flipped at every Monte Carlo step. The size of the cluster is chosen to preserve detailed balance. The Wolff algorithm generates a Boltzman weighted set of spin configurations from where it is possible to calculate canonical thermodynamic averages. This cluster algorithm has previously been used to study systems with inhomogeneous local couplings such as the dilute Ising magnet~\cite{Wang1990} and Sierpinski carpets~\cite{Monceau1998, Monceau2001, Monceau2003, Monceau2002} and the efficiency of the Wolff algorithm seems to increase as the dimension is lowered ~\cite{Monceau2002}. As a result the Wolff algorithm is preferred over Metropolis for the structures in Fig.~\ref{fig:structures}. This is quantified in Section \ref{sec:WolffGood}.

In the limit of large $N$, \Tc\ will correspond to the temperature where the magnetic susceptibility is maximized \Tchimax. To calculate \TchimaxN\ for finite $N$, the histogram method is used. For a specific temperature, \To, the states randomly generated by the Wolff algorithm follow the Boltzmann distribution and can be used to calculate very good estimates of the thermal averages. The histogram method approximates the thermal averages at nearby temperatures by re-weighting the probability of choosing a spin configuration with exponential factors that account for the difference between the temperature of interest and \To. The distance $\Delta T$ which can reliably  be extrapolated away from \To\ is given by~\cite{Newman_book}:

\begin{equation}
\left[\dfrac{\Delta T}{T_{0}}\right]^{2}=\dfrac{1}{Nc\left(T_{0}\right)}.
\label{eq:Trelia}
\end{equation}

\noindent We find that it is safe to extrapolate $\pm 2\Delta T$ from the calculated central temperature, \To. Going two standard deviations away from the mean sample energy still leaves $5\%$ of the samples in the region around the peak of the reweighted histogram. For a collection of a million independent spin configurations, $5\%$ is 50000 samples which yield a reasonable estimate of the internal energy.

Finally, the standard finite size scaling analysis developed by Fisher~\cite{Fisher1972,Newman_book} is used to determine the critical exponents from the behavior of thermodynamic averages as a function of the system size measured in linear dimension, $L$. According to the standard scaling hypothesis, and provided that the size of the system is large enough, the following scaling properties are expected at the critical point: $c\propto L^{\frac{\alpha}{\nu}}$, $m\propto L^{\frac{-\beta}{\nu}}$ and $\chi\propto L^{\frac{\gamma}{\nu}}$, where $\nu$ is the correlation length exponent. The correlation length scales as $\xi(T)\propto\left|t\right|^{-\nu}$, where $t=\left|T-T_{c}\right|/T_{c}$ is a reduced temperature.

For structures with well-defined dimension the linear size follows:

\begin{equation}
L=N^{1/d},
\label{eq:linsz}
\end {equation}
The dimensions of the three structures of Fig.~\ref{fig:structures} are unknown. We assume that Eq. \ref{eq:linsz} holds and define $\nu'=\nu\cdot d$ as a correlation length exponent scaled to the system size.

Finite size effects replace the divergences at the critical point by finite peaks shifted away from \Tc. Effective critical temperatures can thus be defined for each size and each physical quantity concerned (magnetic susceptibility for example) as the positions of these maxima. The shift away from \Tc, to first order approximation can be written as
\begin{equation}
T^{\chi}_{max}=T_c+\chi_{0}\cdot N^{-1/\nu'}
\label{eq:fssfit}
\end {equation}
for the case of susceptibility. A fit of \Tchimax\ against the system size $N$ using Eqn.~\ref{eq:fssfit} gives an estimate of \Tc, \cio\ and \nup.

Provided that \Tc\ and \nup are known with a sufficient accuracy, the following power laws are observed at the critical point: $c(N,T_{c})\propto N^{\frac{\alpha}{\nu'}}$, $m(N,T_{c})\propto N^{-\frac{\beta}{\nu'}}$, and $\chi(N,T_{c})\propto N^{\frac{\gamma}{\nu'}}$. The computation of the critical exponents $\alpha$, $\beta$, $\gamma$ can be deduced from the dependence on size of $c$, $m$ and $\chi$, respectively.

\section{\label{sec:Res}Results and Discussion}
We have calculated \Tc\ for the three structures of Fig.~\ref{fig:structures} from finite-size scaling analysis of the magnetic susceptibility using concatenation levels $k=4$ to $k=7$. The thermodynamic averages were calculated from sets of millions of spin configurations generated by the Wolff algorithm. For each case, the magnetization autocorrelation function was calculated to find the number of successive cluster flips that separate independent spin-configurations. A fit of the auto-correlation function to an exponential reveals that for the structures studied here the autocorrelation time $\tau_{steps}$ is less than a single step ranging from 0.3 to 0.8 cluster flips. This is in contrast to the Metropolis method where initial attempts returned autocorrelation times between $300N$ and $5000N$ possible single-spin flips. Our thermal averages include every-other Wolff Monte-Carlo step after an initial thermalization period of $5\times 10^3$ steps. Once \Tc\ is predicted, the size effects on $c$, $m$, $\chi$, and $\tau_{steps}$ are studied. We find that the thermodynamics of finite structures can be described by critical scaling exponents and that the Wolff algorithm is efficient on these structures at the critical region.

\begin{figure}
\centering
\includegraphics[width=0.5\textwidth]{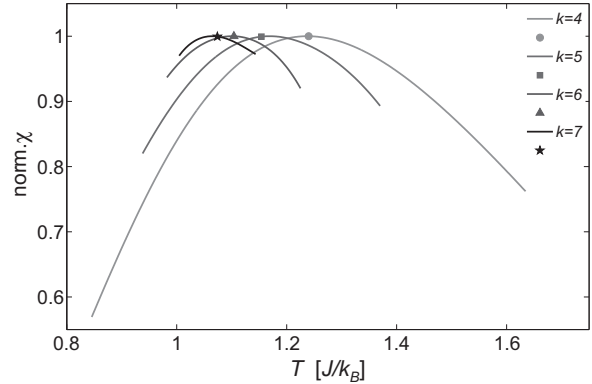}
\caption{\label{fig:hist} Normalized magnetic susceptibilities as a function of $T$ for different concatenation levels predicted using the histogram method (Structure 3). The solid points are temperatures at which Wolff cluster simulations are performed to obtain a set of $1 \times 10^{6}$ uncorrelated samples. Temperature ranges used in predictions are estimated from Eqn.~\ref{eq:Trelia} and shown in Table~\ref{tab:magsus}.}
\end{figure}

\subsection{Finite size effects}
The positions of the effective temperatures, \TchimaxN\, are first estimated by processing the data from short runs of $3 \times 10^{4}$ Wolff Monte Carlo steps from 0.05 to 2 every 0.05 $T$. A second set of short runs are performed over a region of 0.5 with 0.01 $T$ resolution centered at guessed \TchimaxN\ values. Magnetic susceptibility maximums and the corresponding effective temperatures are computed more precisely using the histogram method. Table~\ref{tab:magsus} shows the temperatures at which Wolff cluster simulations are performed to obtain a set of $1 \times 10^{6}$ uncorrelated samples. The magnetic susceptibility is calculated from Eqn.~\ref{eq:magsus} and is re-weighted using the histogram method over the reliable temperature range estimated from Eqn.~\ref{eq:Trelia}. Figure~\ref{fig:hist} shows as an example the results of one of these experimental runs on Structure 3. We repeat this procedure five times for each level of concatenation and structure to check the reliability of the histogram method and to give error estimates on effective temperatures (\Tchimax\ columns of Table~\ref{tab:magsus}). The \Tchimax\ monotonically decreases with the system size for all structures.

\begin{table*}
\caption{\label{tab:magsus} Simulated temperature , $T_{sim}$, (confidence region $2\Delta T$ as per Eqn.~\ref{eq:Trelia}) and the related \Tchimax\ obtained from the Histogram method. The reported temperatures are in units of $J/k_{B}$ and uncertainties quoted are $2\sigma$ errors.}
\begin{ruledtabular}
\begin{tabular}{ccccccccc}
& \multicolumn{4}{c}{$T_{sim}$} & \multicolumn{4}{c}{$T^{\chi}_{max}(N)$} \\
Structure & $k=4$ & $k=5$ & $k=6$ & $k=7$ & $k=4$ & $k=5$ & $k=6$ & $k=7$ \\
\hline
1 & 0.735(251) & 0.685(137) & 0.645(76) & 0.620(43) & 0.736(1) & 0.687(1) & 0.650(5) & 0.611(6) \\
2 & 0.785(322) & 0.725(182) & 0.675(105) & 0.640(60) & 0.783(1) & 0.723(2) & 0.678(3) & 0.646(3) \\
3 & 1.240(394) & 1.155(216) & 1.105(122) & 1.075(70) & 1.239(1) & 1.162(4) & 1.099(6) & 1.056(8) \\
\end{tabular}
\end{ruledtabular}
\end{table*}

\begin{figure}
\centering
\includegraphics[width=0.5\textwidth]{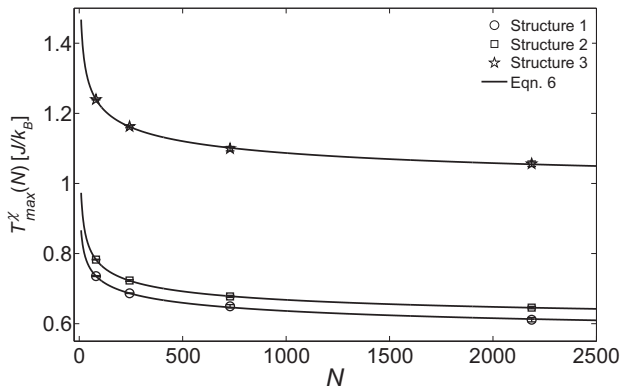}
\caption{\label{fig:fssfit} \Tchimax\ as a function of $N$ for each of the memory stabilizers of Fig~\ref{fig:structures}. The solid lines are fits to the points using Eqn.~\ref{eq:fssfit} (fitting parameters \Tc, \cio\ and \nup\ are reported in Table~\ref{tab:fssfit}).}
\end{figure}

The values of \Tchimax\ as a function of $N$ are plotted in Fig.~\ref{fig:fssfit} for each structure. The solid lines are fits to the points using Eqn.~\ref{eq:fssfit} from which \Tc, \cio\ and \nup\ are obtained (see Table~\ref{tab:fssfit}). Better estimates of these parameters would require additional data points. Unfortunately, for $k<4$, Eqn.~\ref{eq:fssfit} is no longer valid due to higher-order scaling corrections in the small $N$ limit. Additional data points would require calculating the thermodynamic properties at higher levels of concatenation. It may be possible to study bigger systems by using the Wang-Landau algorithm~\cite{Wang2001} with a two dimensional energy and magnetization joint density of states~\cite{Zhou2006}.

The results of the simulations show that one way to increase the critical temperature is by adding generators to each spin site. However, there is not a clear connection between coordination number and \Tc\ (Tables ~\ref{tab:structures} and ~\ref{tab:fssfit}) as the critical temperature does not follow a 1, 1$\frac{1}{3}$, 1$\frac{1}{2}$ progression when going from Structures 1 to 3. When adding generators to go from Structure 1 to Structure 2 (looping only at the $k=1$ level), the increase in \Tc\ is modest. Structure 3 has loops at all concatenation levels, and \Tc\ is almost doubled in comparison to the one of Structure 1. Another distinction of Structure 3 is its higher symmetry. The extra bond changes the energy gap between differing spin configurations and, due to symmetry, changes the underlying density of states at a given energy.

\begin{table}
\caption{\label{tab:fssfit}Finite size scaling law results (\Tc, $\chi_{0}$, $\nu'$) using data from Table~\ref{tab:magsus} in Eqn.~\ref{eq:fssfit} for the three structures shown in Fig.~\ref{fig:structures}. Uncertainties quoted are $2\sigma$ errors.}
\begin{ruledtabular}
\begin{tabular}{cccc}
Structure & \Tc & $\chi_{0}$ & $\nu'$\\ \hline
1 & 0.455(111) & 0.603(35) & 5.747(2.610) \\
2 & 0.552(16) & 0.769(41) & 3.648(367) \\
3 & 0.890(73) & 0.953(69) & 4.374(1.199) \\
\end{tabular}
\end{ruledtabular}
\end{table}

We attempted to study the finite size effects on the heat capacity to further validate \Tc\ predictions. A wide fluctuation of the position of \Tcmax\ from experiment to experiment was observed. It is hard to follow trends from the heights of specific heat peaks for different system sizes and structures. Bhanot \textit{et al.}~\cite{Bhanot1985} pointed out that when space dimensionality is lower than 2, $\alpha$ is expected to be negative, and the specific heat versus temperature peak broadens as the system size increases. We were unable to extract $\alpha/\nu'$ in a reliable way from fits using \Tcmax\ and $c$. Monceau and Perreau encounter similar problems on fractal structures of dimensionality between one and two~\cite{Monceau2001}.

\subsection{Magnetization and magnetic susceptibility at \Tc}
\begin{figure}
\centering
\includegraphics[width=0.5\textwidth]{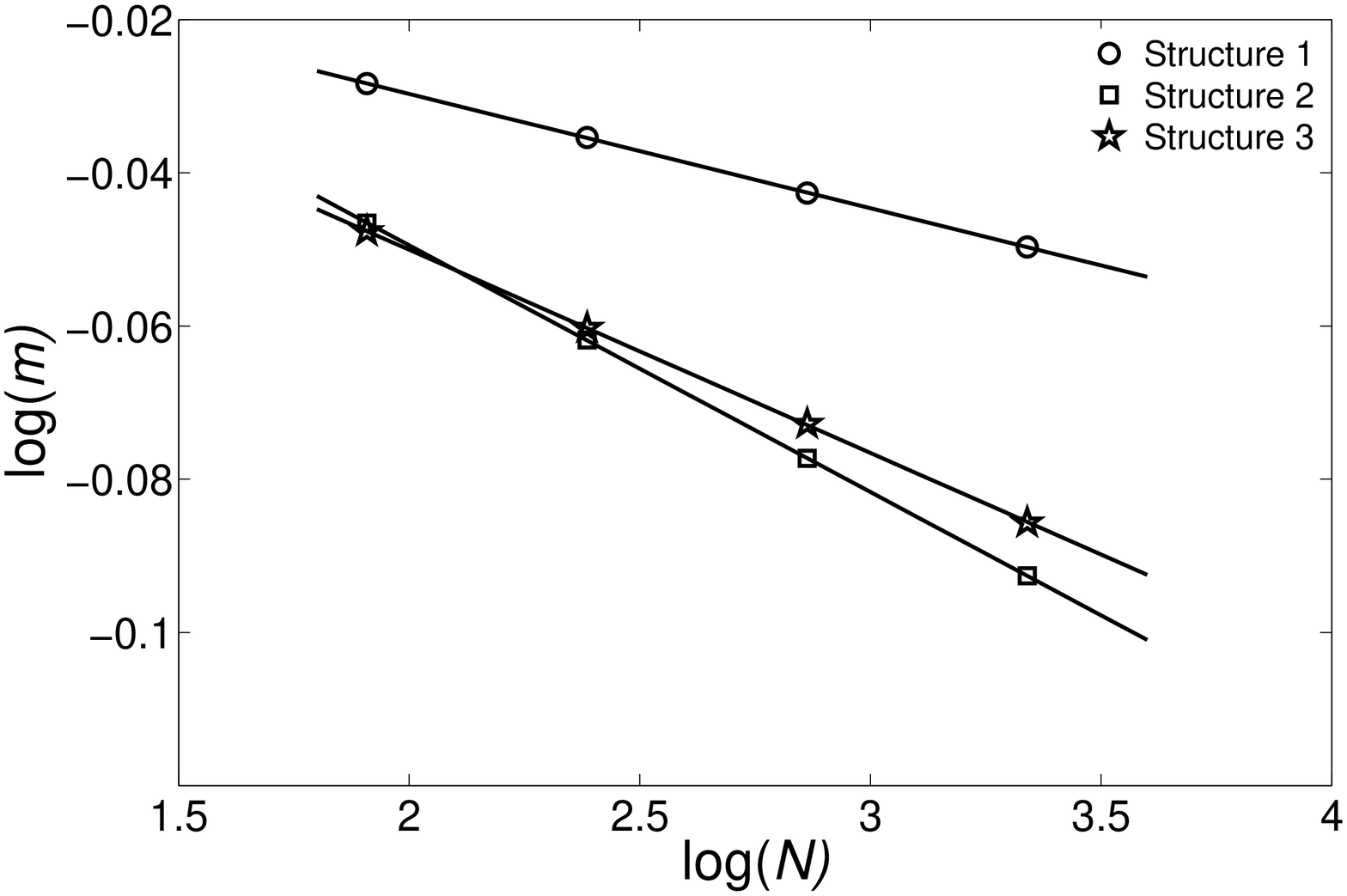}
\caption{\label{fig:pwrlawmag} Magnetizations at critical temperature against system size $N$ for each memory stabilizer on log-log scale. Solid lines show least square fits to power laws from where the $\beta/\nu'$ exponents are calculated.}
\end{figure}

\begin{figure}
\centering
\includegraphics[width=0.5\textwidth]{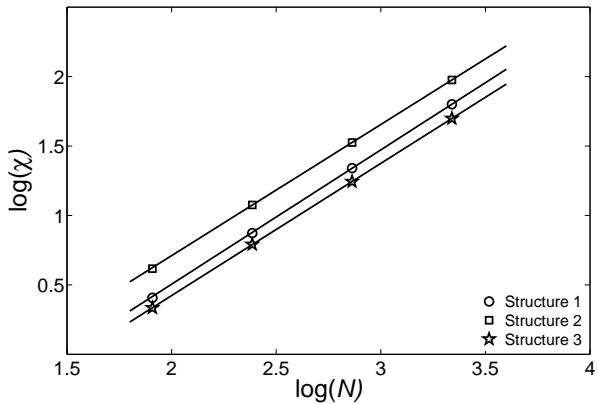}
\caption{\label{fig:pwrlawsus} Magnetic susceptibilities at critical temperature against $N$ for each stabilizer structure on log-log scale. Solid lines show least square fits from where the $\gamma/\nu'$ exponents are calculated.}
\end{figure}

The mean values of the magnetization and the zero-field susceptibility are obtained from simulations at the previously calculated critical temperatures \Tc\ (Table~\ref{tab:fssfit}). We use a single set of $1 \times 10^{6}$ uncorrelated samples for the analysis in this section. The power laws $m(N,T_{c})\propto N^{-\frac{\beta}{\nu'}}$ and $\chi(N,T_{c})\propto N^{\frac{\gamma}{\nu'}}$ are satisfied. Figures~\ref{fig:pwrlawmag} and~\ref{fig:pwrlawsus} show plots of the average absolute magnetization per spin and magnetic susceptibility versus system size $N$ on a log-log scale respectively. The results of the least square analysis are displayed in Table~\ref{tab:pwexp}.

\begin{table}
\caption{\label{tab:pwexp}Exponents obtained from power law behavior of magnetization and susceptibility at critical temperature for each memory stabilizer structure. Uncertainties quoted are $2\sigma$ errors.}
\begin{ruledtabular}
\begin{tabular}{ccc}
Structure & $\beta/\nu'$ & $\gamma/\nu'$ \\ \hline
1 & 0.015 & 0.967(7) \\
2 & 0.032 & 0.944(2) \\
3 & 0.027 & 0.952(2) \\
\end{tabular}
\end{ruledtabular}
\end{table}

The exponent $\alpha$ is deduced from the Rushbrooke scaling law $\alpha = 2 - 2\beta - \gamma$ and it takes, as expected, negative values for the three memory stabilizers. Table~\ref{tab:criexp} shows the set of critical exponents ($\alpha$, $\beta$, $\gamma$) for each structure. We write the Rushbrooke and Josephson scaling law $d = \gamma/\nu + 2\beta/\nu$ as a function of \nup\ to get rid of the unknown dimension $d$. The last column of Table~\ref{tab:criexp} shows that the scaling law is satisfied within an error of less than $1\%$. This means that the magnetization (the order parameter of interest for memory preservation) is continuous at \Tc. The uncertainty in the critical exponents is quite big for the three graphs. It is not possible to conlcude whether or not the memory structures share the same set of critical exponents. All three structures may be in some weak universality class in which critical exponents may not only depend upon the symmetry of order parameters and fractal dimensions, but also upon their geometry. This seems to be the case for Ising magnets from Sierpinski fractals of non-integer dimensions between one and three~\cite{Monceau1998, Monceau2001, Monceau2003,Monceau2002}.

\begin{table}
\caption{\label{tab:criexp}Set of critical exponents ($\alpha$, $\beta$, $\gamma$) of three structures shown in Fig.~\ref{fig:structures}. Last column checks consistency of the results by using the Rushbrooke and Josephson's scaling laws as discussed in the text. Uncertainties quoted are $2\sigma$ errors.}
\begin{ruledtabular}
\begin{tabular}{ccccc}
Structure & $\alpha$ & $\beta$ & $\gamma$ & $1=\dfrac{\gamma}{\nu'}-2\dfrac{\beta}{\nu'}$\\ \hline
1 & -3.730(2.526) & 0.086(39) & 5.558(2.525) & 0.997(3) \\
2 & -1.679(347) & 0.117(12) & 3.444(347) & 1.009(1) \\
3 & -2.395(1.143) & 0.116(32) & 4.163(1.141) & 1.005(1) \\
\end{tabular}
\end{ruledtabular}
\end{table}

\subsection{Wolff algorithm efficiency at \Tc}\label{sec:WolffGood}
\begin{figure}
\centering
\includegraphics[width=0.5\textwidth]{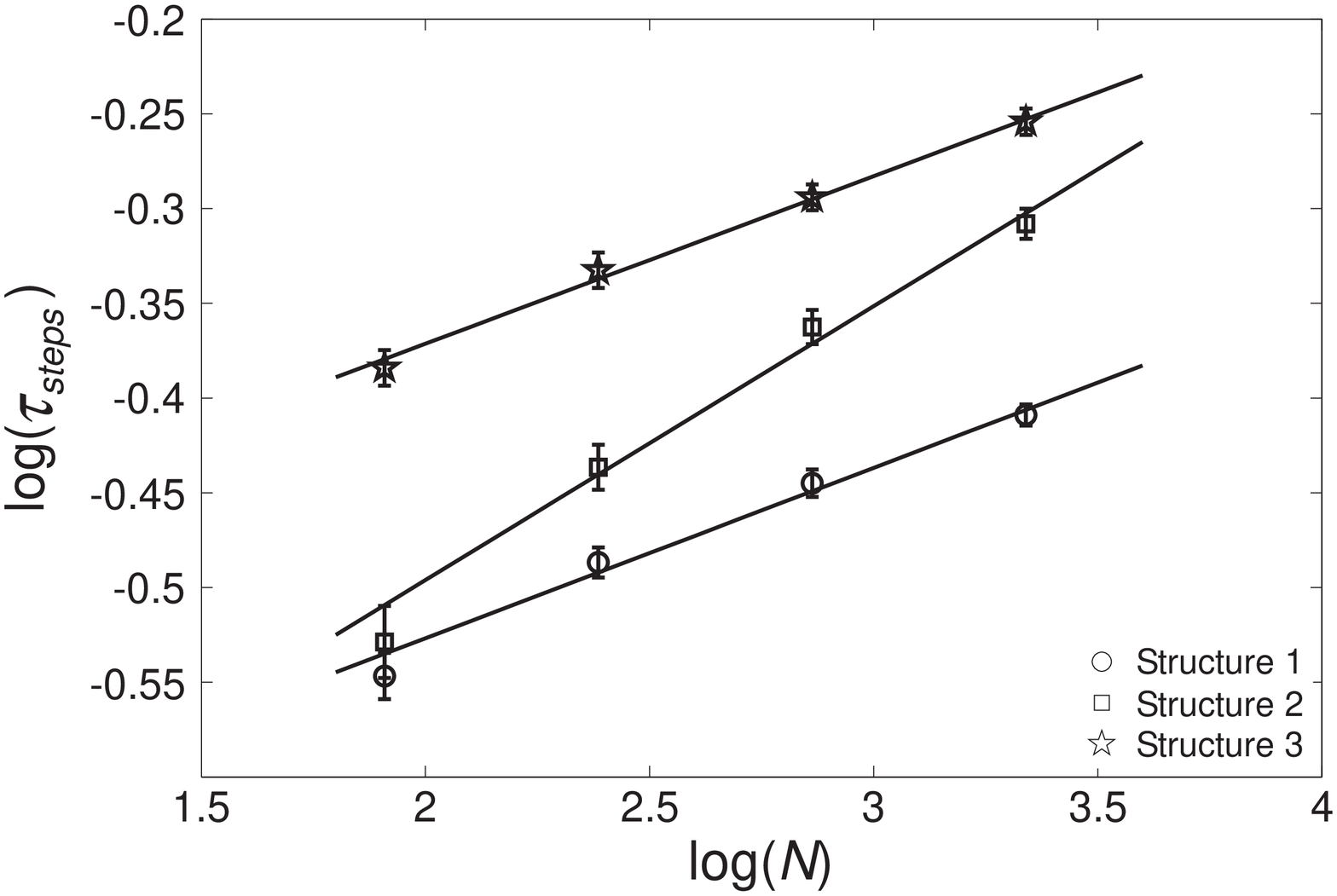}
\caption{\label{fig:tauWce} Magnetization Wolff autocorrelation times (in number of cluster flips) against system size $N$ for each memory stabilizer structure on log-log scale. The solid lines are fits to $\tau_{steps} \propto N^{z_{0}/d}$ from where measured critical exponents $z_{0}/d$ are calculated.}
\end{figure}

The dynamical aspects of the Wolff algorithm when applied to memory stabilizers are analyzed. We take five runs of $5 \times 10^{5}$ cluster flips at each concatenation level $k=4-7$ and for each structure to calculate the magnetization autocorrelation function. The autocorrelations are fit to a single exponential decay to obtain Wolff autocorrelation times $\tau_{steps}$ (see Table~\ref{tab:tauWce}). As shown in Fig.~\ref{fig:tauWce}, magnetization autocorrelation times follow the power law $\tau_{steps} \propto N^{z_{0}/d}$ at the critical temperature. The Wolff dynamical critical exponent, $z/d$, associated with memory stabilizer structures is defined as

\begin{equation}
\frac{z}{d}=\frac{z_{0}}{d}+\frac{\gamma}{\nu'}-1.
\label{eq:Wdynce}
\end {equation}

The Wolff algorithm is very efficient in reducing the critical slowing down (increase in correlation time as \Tc\ is approached) for the stabilizer structures of Fig.~\ref{fig:structures}. The dynamical critical exponents are very low compared to the Metropolis or Wolff algorithm on the 2D Ising Model, where $z/d=1.0835$ and $z/d=0.125$ respectively~\cite{Newman_book}.

\begin{table}
\caption{\label{tab:tauWce}Wolff autocorrelation times, $\tau_{steps}$, measured from the decay of the magnetization autocorrelation function for each memory stabilizer structure and system size. Last two rows show the measured critical exponent $z_{0}/d$, obtained from power law fits, and the Wolff dynamical critical exponent $z/d$ (from Eqn.~\ref{eq:Wdynce}). Uncertainties quoted are $2\sigma$ errors.}
\begin{ruledtabular}
\begin{tabular}{cccc}
\multicolumn{1}{c}{} & \multicolumn{3}{c}{Structure} \\
$k$ & 1 & 2 & 3 \\ \hline
4 & 0.284(16) & 0.296(26) & 0.413(18) \\
5 & 0.326(12) & 0.366(20) & 0.465(19) \\
6 & 0.359(11) & 0.434(18) & 0.508(16) \\
7 & 0.390(10) & 0.492(18) & 0.557(18) \\
$z_{0}/d$ & 0.090(15) & 0.145(26) & 0.088(8) \\
$z/d$ & 0.057(16) & 0.089(26) & 0.040(8) \\
\end{tabular}
\end{ruledtabular}
\end{table}

\section{\label{sec:Con}Conclusions}
We have used the Wolff algorithm together with the histogram method and finite-size scaling analysis to calculate critical temperatures of Hamiltonians based on concatenated error-correction codes. The three simple two-body-interaction structures investigated have different levels of connectivity. We find that the relationship between coordination number and critical temperature is not obvious. The intriguing result is that the number of generators is less important than the structure. For a minimum number of generators, one can have either a linear Ising model with no phase transition or Structure 1 with a finite phase transition. If one adds additional connections or generators, the results can range from a modest increase in \Tc\ (Structure 1 and Structure 2) to a doubling of \Tc\ (Structure 1 and Structure 3). Whether these insights can be applied to self-correcting quantum systems is an open question.

Scaling properties of the magnetization and magnetic susceptibility satisfy power-law fits as function of total number of spins $N$. Each structure exhibits second order or continuous phase transition. We report the set of critical exponents ($\alpha$, $\beta$, $\gamma$), and by fitting the decay of the magnetization autocorrelation functions at the critical points we calculate Wolff dynamical critical exponents. It is possible that all three structures are in some weak universality class but the current study does not show this.

For quantum information, a thermodynamically unstable memory that has a large kinetic barrier could be useful for preserving information. It is possible that many of the stabilizer codes that are not self-correcting memories could satisfy this relaxed condition. Although the kinetics depends strongly on the details of the specific system-bath coupling~\cite{Brown2007}, the work here suggests that the choice of geometry and generators could lead to large differences in the effective information preservation. 

\begin{acknowledgments}
Authors would like to thank Dave Bacon and Rigoberto Hernandez for useful discussions. Some of the simulations were performed at the facilities of the Center for Computational Molecular Science and Technology (CCMST) and ECE Academic Labs. This work was supported by the Georgia Institute of Technology.
\end{acknowledgments}

\newpage %Just because of unusual number of tables stacked at end

\end{document}